\documentclass[fleqn,usenatbib]{mnras}

\usepackage{newtxtext,newtxmath}

\usepackage[T1]{fontenc}

\DeclareRobustCommand{\VAN}[3]{#2}
\let\VANthebibliography\thebibliography
\def\thebibliography{\DeclareRobustCommand{\VAN}[3]{##3}\VANthebibliography}

\usepackage{graphicx}	
\usepackage{amsmath}	
\usepackage{comment}
\usepackage{subfig}

\title[Magnetic He WD in NGC~6397]{A candidate magnetic helium core white dwarf in the globular cluster NGC~6397}

\author[Manuel Pichardo Marcano et al.]{Manuel Pichardo Marcano$^{1,2}$ \thanks{Contact e-mail: \href{mpichardomarcano@amnh.org}{mpichardomarcano@amnh.org}}{Liliana E. Rivera Sandoval$^{3}$, Thomas J.  Maccarone $^{1}$, Rene D. Rohrmann$^{4}$, }
\newauthor{Craig O. Heinke$^{5}$, Diogo Belloni$^{6}$, Leandro G. Althaus$^{7,8}$, Arash Bahramian$^{9}$}
\\
$^{1}$Department of Physics and Astronomy, Texas Tech University, Lubbock, TX 79409\\
$^{2}$Department of Astrophysics, American Museum of Natural History, New York, NY, USA\\
$^{3}$Department of Physics and Astronomy, University of Texas Rio Grande Valley, Brownsville, TX
78520, USA\\
$^{4}$Instituto de Ciencias Astronómicas, de la Tierra y del Espacio (CONICET-UNSJ), Av. España 1512 (sur), 5400 San Juan, Argentina\\
$^{5}$Department of Physics, University of Alberta, CCIS 4-183, Edmonton, AB, T6G 2E1, Canada\\
$^{6}$Departamento de F\'isica, Universidad T\'ecnica Federico Santa Mar\'ia, Av. España 1680, Valpara\'iso, Chile\\
$^{7}$Grupo de Evoluci\'on Estelar y Pulsaciones. Facultad de Ciencias Astronómicas y Geofísicas,Universidad Nacional de La Plata\\, Paseo del Bosque s/n, 1900 La Plata, Argentina\\
$^{8}$IALP - CONICET, La Plata, Argentina\\
$^{9}$International Centre for Radio Astronomy Research Curtin University, GPO Box U1987, Perth, WA 6845, Australia
}

\date{Accepted XXX. Received YYY; in original form ZZZ}

\pubyear{2023}

\begin{document}
\label{firstpage}
\pagerange{\pageref{firstpage}--\pageref{lastpage}}
\maketitle

\begin{abstract}
We report a peculiar variable blue star in the globular cluster NGC~6397, using Hubble Space Telescope optical imaging. Its position in the colour-magnitude diagrams, and its spectrum, are consistent with this star being a helium core white dwarf (He WD) in a binary system. The optical light curve shows a periodicity at $18.5$ hours. We argue that this periodicity is due to the rotation  of the WD and possibly due to magnetic spots on the surface of the WD. This would make this object the first candidate magnetic He WD in any globular cluster (GC), and the first candidate magnetic WD in a detached binary system in any GC and one of the few He WDs with a known rotation period and of magnetic nature. Another possibility is that this system is a He WD in a binary system with another WD or another degenerate object, 
which 
would make this object one of the few candidate non-accreting double degenerate binaries in any GC. 
\end{abstract}




\section{Introduction}

NGC 6397 is the closest (2.5 kpc) core-collapsed Galactic globular cluster (GC). Due to its proximity and low reddening, $E(B - V) = 0.18$ \citep{harris_catalog_1996,mclaughlin_resolved_2005,BaumgardtDistance2021}, NGC 6397 has been extensively studied at different wavelengths. One particular population that has been studied is the white dwarfs (WDs), in large part due to the capabilities of the Hubble Space Telescope (HST),  especially its angular resolution which makes it capable of studying sources in crowded regions like GCs \citep[e.g.][]{CoolWDSequence1996ApJ...468..655C,HansenWDSequence2007ApJ...671..380H,Richer2006Sci...313..936R,Richer2008AJ....135.2141R}. HST has allowed the study of the cooling sequence of isolated carbon-oxygen WDs (CO WDs) to determine, among other things, the age and distance of the cluster \citep{HansenWDSequence2007ApJ...671..380H,Richer2008AJ....135.2141R}.


Besides CO WDs, a more exotic and less studied population of WDs with helium cores (He WDs) is known to exist in NGC~6397 \citep{Cool98,edmonds_cataclysmic_1999}.
He WDs are not expected to be the typical outcome of single stellar evolution because the Universe is not old enough for the very low mass stars (i.e. stars of too low a mass to fuse helium into carbon) to finish their lifetimes. The He WDs thus must be produced from stars that have had their evolution halted during binary evolution.
In the Milky Way field, they are predominantly found in binary systems \cite[$\gtrsim70$~per~cent, e.g.][]{Marsh1995MNRAS.275..828M,BrownBinaryHeWDs2011ApJ...730...67B}, and in particular {\it all} extremely-low-mass He WDs ($\lesssim0.2~{\rm M}_\odot$) are known to be members of binaries \citep[e.g.][]{Brown_2020}.
Characterizing 
the putative binary population of He WDs is important as these objects are potential sources of low-frequency gravitational waves \citep[e.g.][]{Nelemans_2013,Kremer2021ApJ...917...28K} as well as progenitors/members of AM~CVns or ultra-compact X-ray binaries, in which a more massive compact object accretes mass from a Roche-lobe filling He WD \citep[e.g.][]{Nelemans_2010}.

Except 
for the 
enhanced mass-loss scenario during single star evolution \citep[e.g.][]{Castellani_1993}, all other formation channels proposed for the origin of single He WDs involve a companion, such as mass ejection by massive planets \citep[e.g.][]{Nelemans_1998}, type Ia supernova stripping \citep[e.g.][]{Justham_2009}, double He WD mergers \citep[e.g.][]{Saio_2000}, and dynamically unstable cataclysmic variable merger  \citep[e.g.][]{Zorotovic_2017}.


 In NGC~6397, three He WD candidates were first identified by \cite{Cool98} as UV stars significantly bluer than the main sequence (MS) but lacking the flickering generally seen in cataclysmic variables (CVs). He WDs lie above the CO WD sequence, due to their unusually large radii (a direct consequence of their low masses, combined with the degenerate nature of WDs), in the gap region between the main sequence (MS) and the CO WD sequence. In the $H\alpha$-R colour-magnitude diagram (CMD), they are redder than the MS showing strong $H\alpha$ absorption, in contrast to CVs which generally show H$\alpha$ excesses. One candidate was later confirmed via spectroscopy by \cite{edmonds_cataclysmic_1999} to be a low-mass He WD. Further studies by \cite{taylor_helium_2001}, using photometry from HST's narrow  $H\alpha$ filter, found 9 additional sources deficient in $H\alpha$ and thus good He WD candidates. Currently, there are 24 good candidate He WDs in NGC~6397 found by \cite{Strickler2009ApJ...699...40S} based on their position in the CMDs. \citealt{Strickler2009ApJ...699...40S} determined photometric masses in the range of $0.2-0.3 M_\odot$, and based on the locations of the objects within the cluster, determined dynamical masses similar to those of the blue stragglers. The difference in the masses argues that these must typically be in binaries with companions that are rather heavy for globular cluster objects; given that the colours of these objects are not affected by their companions, the companions are likely to be other compact objects.
 

Here we present the optical counterpart and light curve of a candidate binary He core WD in NGC~6397  that shows a periodicity of $18.5$ hours. We interpret this periodicity in the light curve as the rotation period of the WD and hence as evidence for magnetism in this system. This would make this object the first magnetic candidate He core WD in any star cluster, and one of the few He WDs with both a known rotation period, and a magnetic nature. 


\section{Data Analysis and Results}

\subsection{Data}

This object was found as a 
variable and blue 
object 
in 
the Survey for Compact Objects and Variable Stars (SCOVaS) (Pichardo Marcano et al., in prep). SCOVaS is a survey for cataclysmic variables and compact binaries in Galactic globular clusters using multiwavelength archival data from HST. The details of the survey will be published in a future paper. 


For NGC 6397 we use  Wide Field and Planetary Camera 2 (WFPC2) data from the parallel field that observed the core of the cluster from the HST large program GO-10424 (PI H. Richer), \cite{Richer2006Sci...313..936R}. The data set consists of 126 orbits,
where each orbit is divided into three exposures in three different filters (F814W, F606W, and F336W). For this work, we use the 126 individual exposures in the filter F336W, with exposure times ranging from 500-700 seconds, taken between mid-March and April 2005 (2005-03-13 to 2005-04-08). The minimum separation between consecutive data points is 74 minutes, the maximum separation between consecutive data points is 3.2 days, and the total baseline is 26 days.

We also use the GO-10257 dataset (PI: Anderson), which provides Advanced Camera for Surveys, Wide Field Channel (ACS/WFC) imaging of the central region of NGC~6397 in the filters F435W ($B$), F625W ($R$), and F658N ($H\alpha$). 

For the photometry of both datasets, we used the software package DOLPHOT \citep{2000PASP..112.1383D}. For the WFPC2 data, we supply DOLPHOT with the calibrated single-exposure image data (c0m) WFPC2 files and the drizzled .drz image as the reference frame for alignment, and for the ACS/WFC we supplied the CTE-corrected  .flc images and a .drz image as reference frames. The final output for both runs from the software lists the position of each star relative to the reference image, together with the measured aperture-corrected magnitudes calibrated to the Vega system for the individual exposures, along with some diagnostic values. We limit the data to measurements containing an error flag of zero meaning that the star was recovered extremely well in the image without contamination due to cosmic rays.

We also made use of the \emph{HST} UV Globular Cluster Survey (HUGS) \citep{Piotto2015AJ,Nardiello2018MNRAS} (programmes GO-10775, GO-13297 ) and the Hubble Space Telescope Atlases of Cluster Kinematics \citep[HACKS,][]{HACKSLibralato2022} (programme GO-10257). The HUGS catalog\footnote{https://archive.stsci.edu/prepds/hugs/} provides photometry in 5 different bands (F275W, F336W, F438W, F606W, and F814W). The HACKS dataset also provides a photometric (F435W, F606W, F625W, F814W, F336W, F438W, F606W) and  proper motion catalogue for the GC NGC~6397\footnote{http://dx.doi.org/10.17909/jpfd-2m08}. We used both public catalogues to build CMDs of the cluster to confirm that our photometric results are consistent with previous studies, and to confirm the membership of our sources. 

We used publicly released spectra from the Multi Unit Spectroscopic Explorer \citep[MUSE][]{MUSECommissioning}. MUSE is an optical  integral-field spectrograph operating in the visible wavelength range at the Very Large Telescope. We used the data from NGC~6397\footnote{http://muse-vlt.eu/science/globular-cluster-ngc-6397/} as published in \cite{Husser2016MUSE} and \cite{kamann_muse_2016}. The data consist of short $\leq 60 s$ exposures to avoid saturation of the brightest cluster giants.



\subsection{Photometic Results}

\begin{figure}
\includegraphics[width=1.0\columnwidth]{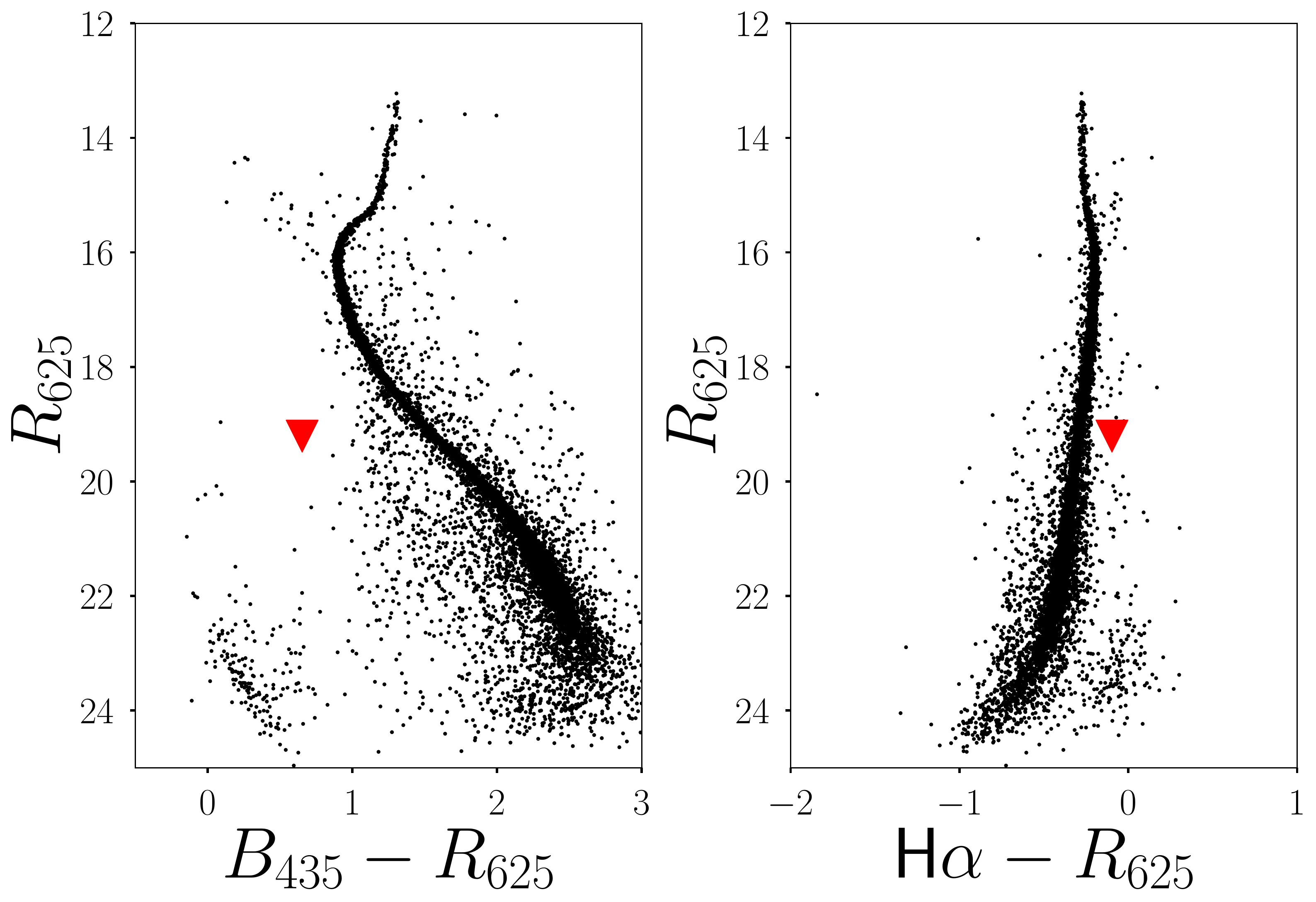}
\caption{HACKS colour-magnitude diagrams (CMDs) for the central regions of NGC~6397. The red triangle shows the position in each CMD of the candidate magnetic helium-core white dwarf.}
\label{fig:Hacks_CMDs}
\end{figure}

\begin{figure*}\captionsetup[subfigure]{labelformat=empty}
  \centering
  \subfloat[][]{\includegraphics[width=1.0\columnwidth]{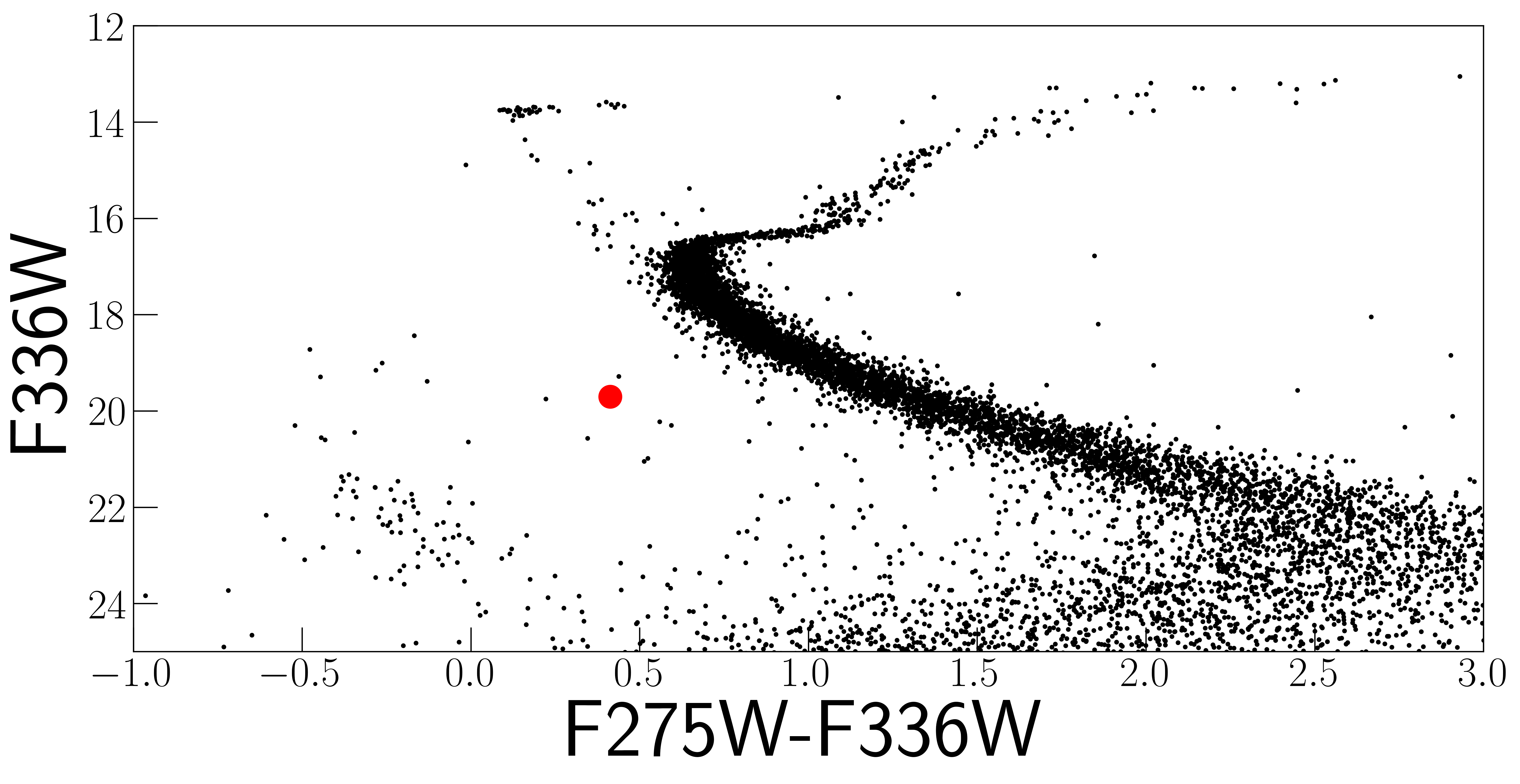} \label{fig:F336WCMD}} 
  \subfloat[][]{\includegraphics[width=1.0\columnwidth]{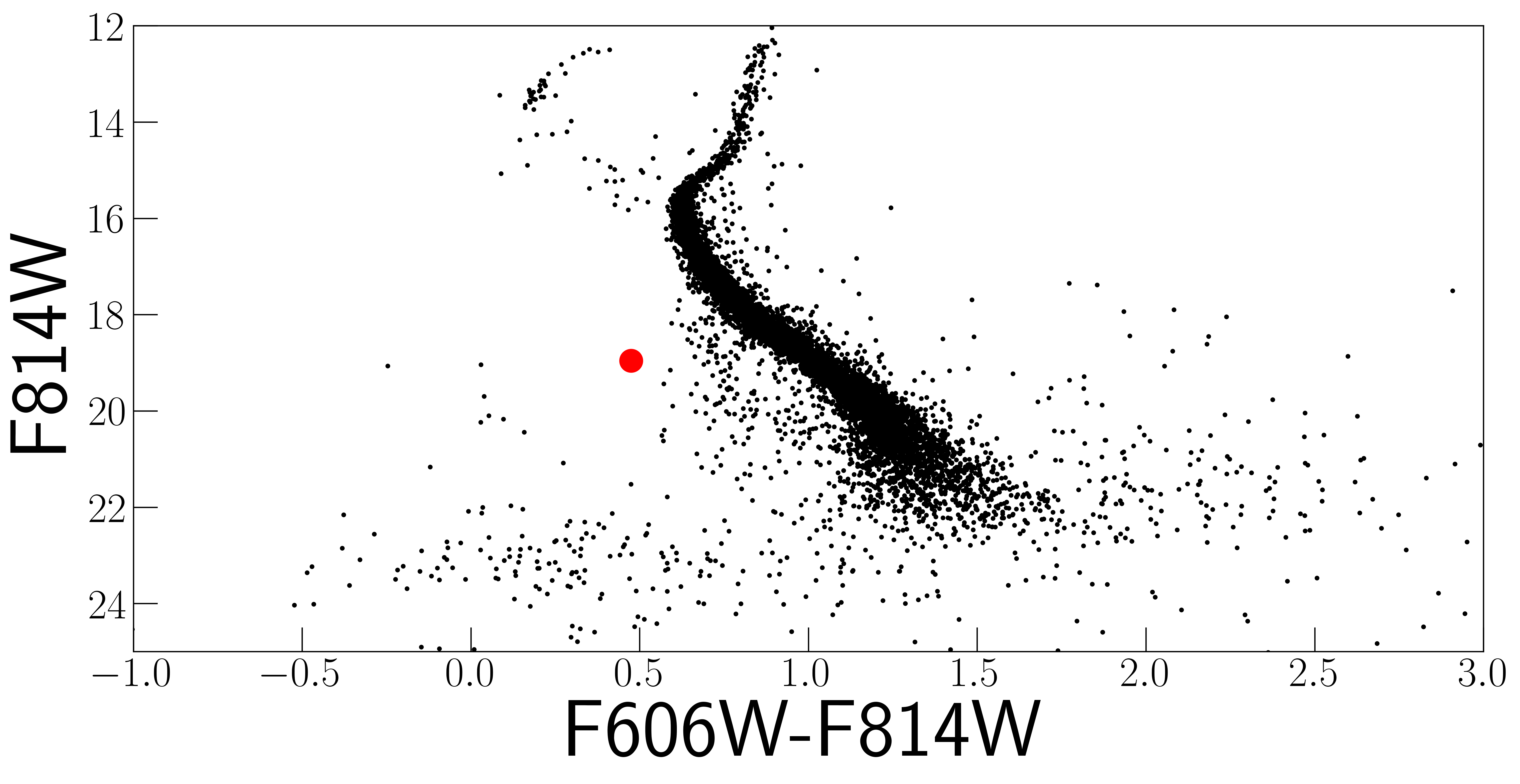} \label{fig:F814WCMD}}
  \caption{CMDs of stars from the HUGS catalogue. The red point indicates the position of the candidate magnetic helium-core white dwarf.} 
  \label{fig:CMDs}

\end{figure*}

The CMDs in figures~\ref{fig:Hacks_CMDs} and \ref{fig:F336WCMD} identify the optical counterpart to the peculiar variable blue star as a red point. 
The left panel of figure~\ref{fig:Hacks_CMDs} shows the source in the gap region between the CO WD sequence and the main sequence. 
The position of the source on the CMDs, and its proper motion (matching the cluster value), are consistent with it being a He WD.

\begin{figure}
\includegraphics[width=1.0\columnwidth]{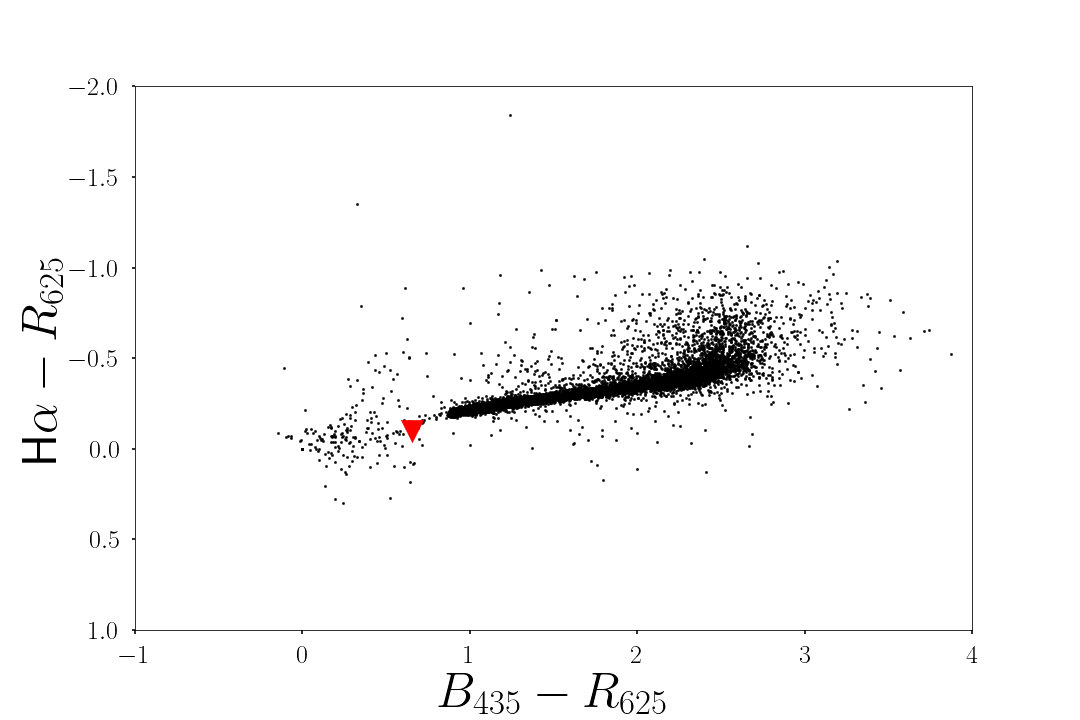}
\caption{HACKS colour-colour diagram for the central regions of NGC~6397. The red triangle shows the position in the CMD of the candidate magnetic helium-core white dwarf.}
\label{fig:ColorColor}
\end{figure}

\begin{figure}
\includegraphics[width=1.0\columnwidth]{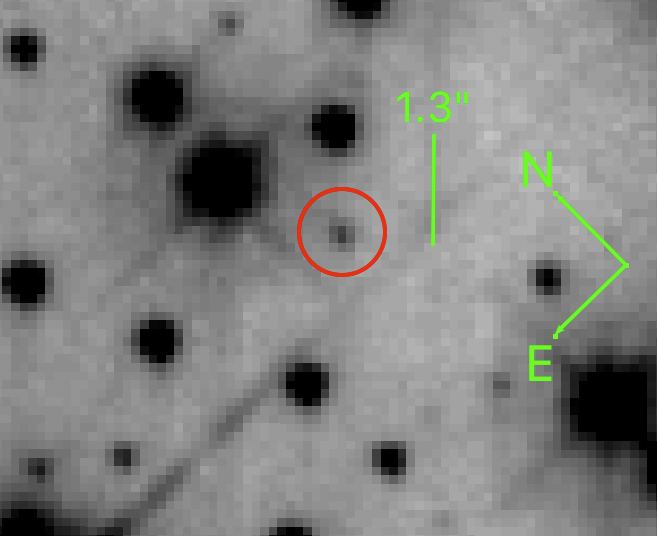}
\caption{F336W image centered at the candidate magnetic He WD. The red circle is 0.5" in radius and centred at 17:40:37.07, -53:40:08.99 (J2000, epoch 2015.0). The position is from the HUGS catalogue {\protect\citep{Piotto2015AJ,Nardiello2018MNRAS}}. }
\label{fig:FindingChart}
\end{figure}

The right panel of figure~\ref{fig:Hacks_CMDs} shows the source to the right of the main sequence, consistent with a stronger $H\alpha$ absorption line. However, as this object is bluer than main sequence stars of the same brightness, a more appropriate test is to construct a color-color diagram, contrasting its $H\alpha$-$R$ and $B$-$R$ colors. Fig. \ref{fig:ColorColor} shows that in such a diagram, this source does not show evidence for  $H\alpha$ emission or greater-than-average absorption.
 A finding chart for this source is shown in fig.~\ref{fig:FindingChart}.

\subsection{MUSE Spectrum}

\begin{figure*}
\includegraphics[width=0.9\textwidth]{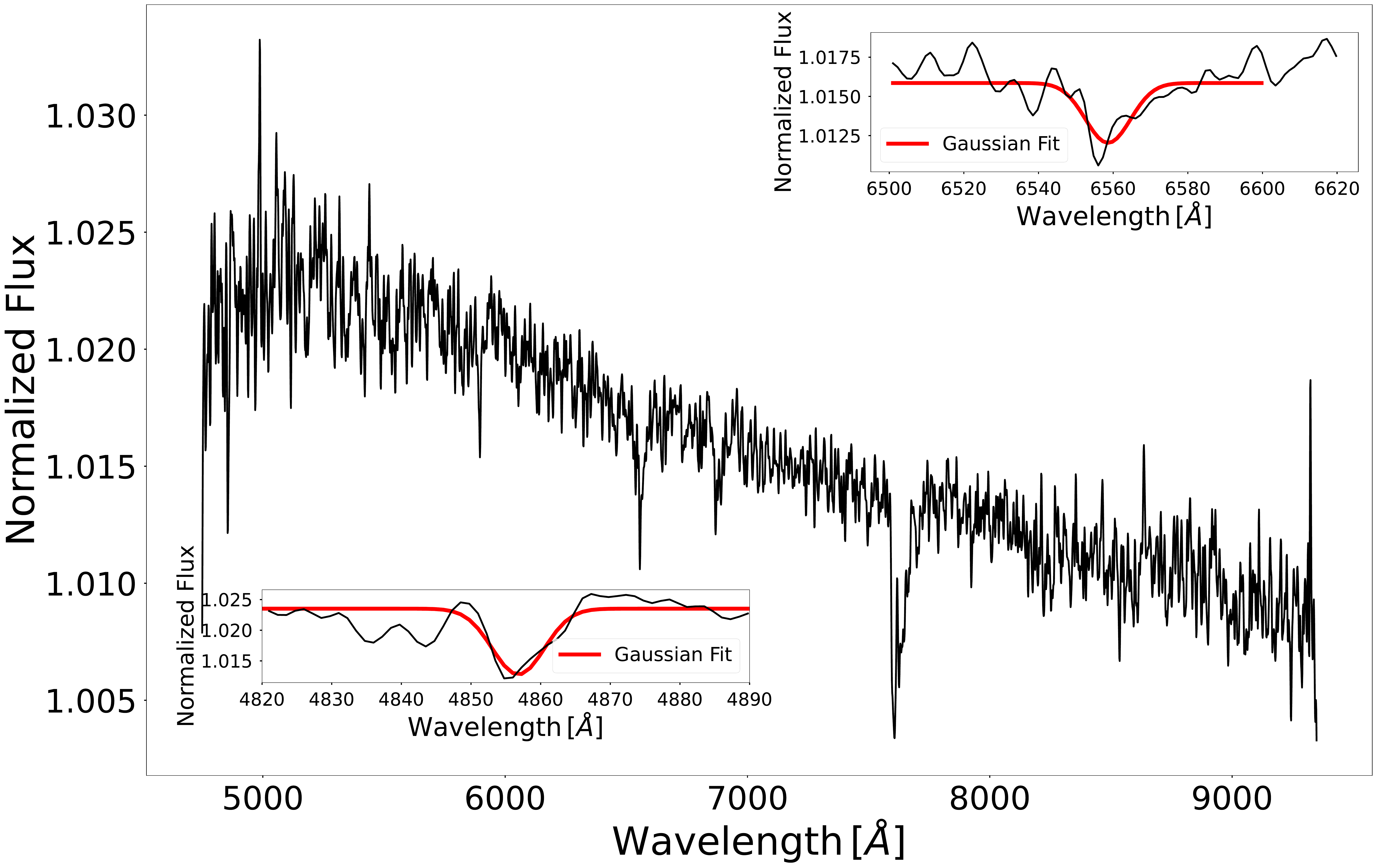}
\caption{Observed MUSE spectrum for the candidate magnetic He WD. The spectrum clearly shows the H$\alpha$ line in absorption. The insets to the upper right, and lower left, show the best fit to the H$\alpha$ and H$\beta$ absorption lines, respectively. 
The centroids of both lines are blue-shifted, showing evidence that the WD is part of a binary system.   }
\label{fig:MUSESpectrumFitBoth}
\end{figure*}

This source is included in the public data release\footnote{http://muse-vlt.eu/science/globular-cluster-ngc-6397/} of MUSE sources (Star ID: 11691 ) as published in \cite{Husser2016MUSE} and \cite{kamann_muse_2016}. The data reduction of the released public spectra was done using the official MUSE pipeline (in versions 0.18.1, 0.92, and 1.0;  \citealt{2012SPIE.8451E..0BW}). The spectrum has a signal-to-noise ratio of 5.9 and an exposure time of 25 seconds. We smooth the spectrum by convolving it with a Gaussian with a standard deviation of 1.5 Angstroms. The smoothed spectrum for this source is shown in figure~\ref{fig:MUSESpectrumFitBoth} along with the Gaussian fit to the $H\alpha$ and $H\beta$ lines as insets. We fitted the Gaussians using the astropy \citep{astropy:2013,astropy:2018} built-in routine after 
detrending 
the spectrum by fitting a polynomial to the continuum, excluding the fitted line. We used the Levenberg-Marquardt algorithm to find the best fit and estimated the errors following \citep{1992PASP..104.1104L}. 
The spectrum shows the H$\alpha$ and H$\beta$ lines in absorption. The best Gaussian fit to the H$\alpha$ line shown in the inset has a mean of 6558.63 \AA \,and a standard deviation of  6.03 \AA. Using the single available spectrum, we measure a radial velocity for the H$\alpha$ line of $-200 \pm 22 $ km/s.  This value is 
reasonably close to 
the radial velocity obtained using the H$\beta$ line of $-282.2 \pm 23 $ km/s. To check the wavelength calibration, we used the atmospheric ozone 
absorption bands centred at  6870 \AA\ \, and 7605 \AA. For their centres we get $7605.10 \pm 0.01$ \AA \, and $6868.83 \pm 0.01$ \AA. For NGC 6397 \cite{Husser2016MUSE} reported a mean radial velocity for the cluster population of $17.8$ km/s with a dispersion of 5.0 km/s. 
Our 
large radial velocity, compared to the mean radial velocity of the members, implies that this source is either a non-member or in a binary. 

\subsection{Light curve and timing analysis}
\label{sec:lc}

\begin{figure}
\includegraphics[width=1.0\columnwidth]{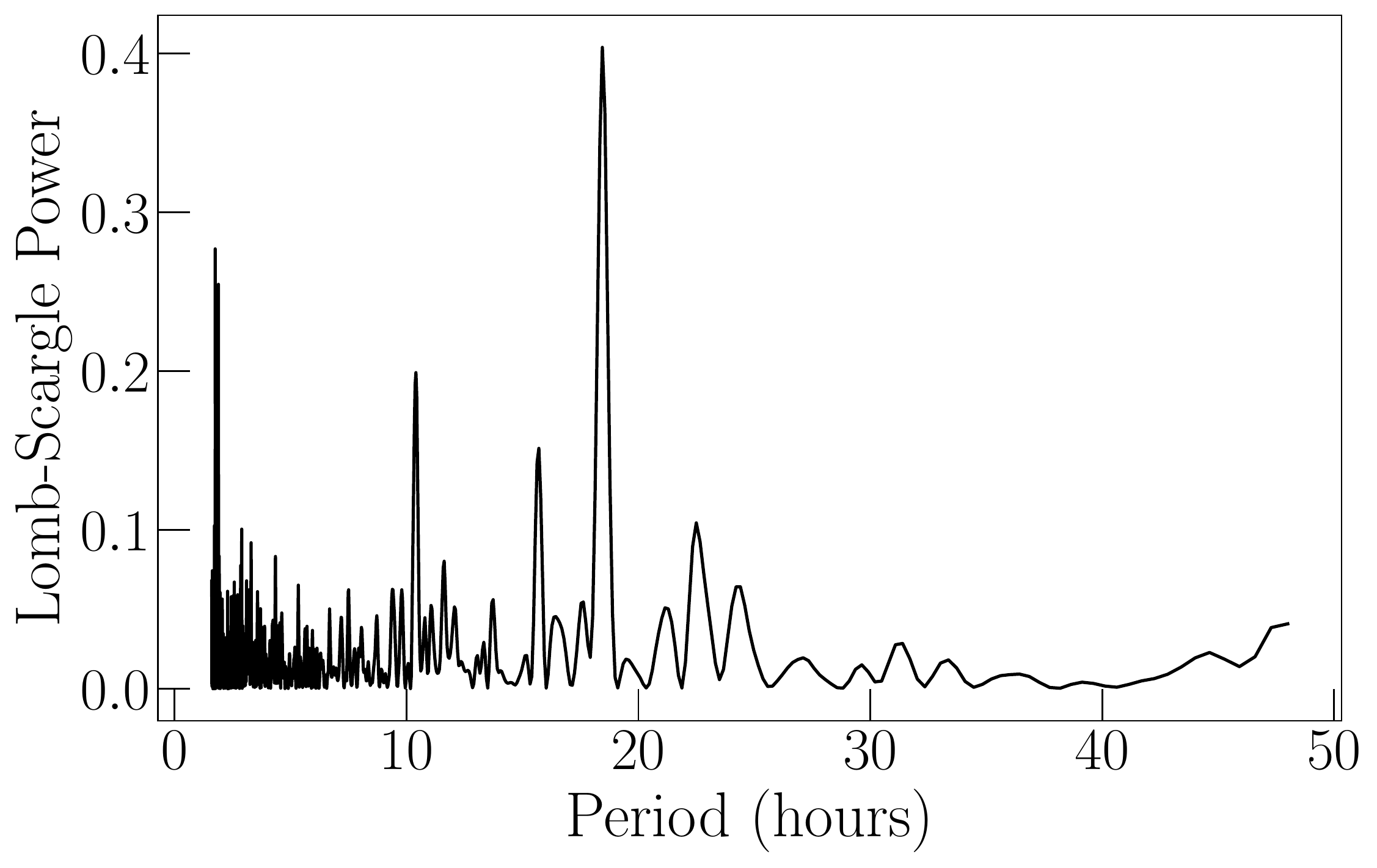}
\caption{Lomb-Scargle periodogram for the candidate magnetic helium-core white dwarf. The periodogram shows an isolated peak at 18.4 hours or 0.769 days. The false-alarm probability of the peak is calculated to be $1.4\times 10^{-8}$ using the method described in {\protect\cite{Baluev}}.}
\label{fig:MagneticHeWDLS}
\end{figure}

\begin{figure}
\includegraphics[width=1.0\columnwidth]{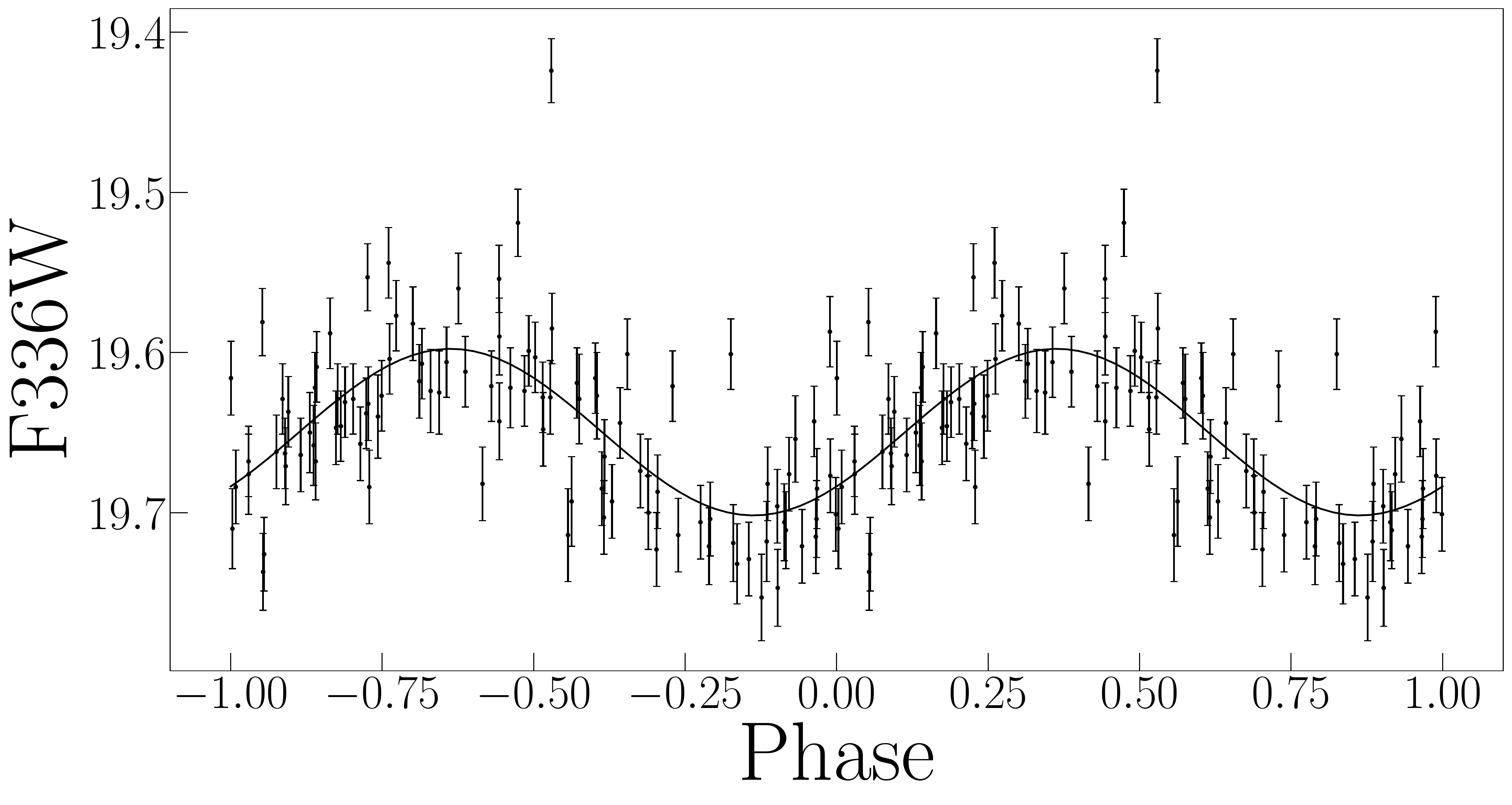}
\caption{Light curve (F336W filter) of the candidate magnetic helium-core white dwarf folded at a period of 18.4 hours or 0.769 days.}
\label{fig:F336WLC Folded}
\end{figure}

The HST F336W light curve shows clear variability of the order of $0.1$ magnitudes. We performed a periodicity search on the light curve using the Lomb-Scargle method \citep{Lomb1976Ap&SS..39..447L,Scargle1982ApJ...263..835S}. The resulting periodogram is shown in figure~\ref{fig:MagneticHeWDLS}. The periodogram shows a significant peak with a false-alarm probability of  $1.4 \times 10^{-8}$, using the method described in \cite{Baluev}, which allows getting an upper limit that is valid for alias-free periodograms and in the absence of strong aperiodic variablity. The peak is at $0.769 \pm 0.01 $ days or  $18.46 \pm 0.24$ hours. The uncertainty on the period was estimated by the standard deviation of the best-fitted Gaussian to the peak. The light curve  folded at $0.769$ days is shown in figure~\ref{fig:F336WLC Folded}. Given that the data set spans about 35 cycles of the period, and that the sampling is fairly dense, we can be confident that this is a real period and not due to red noise.

\begin{table*}
\centering

\resizebox{\linewidth}{!}{%
\begin{tabular}{|l|l|l|l|l|l|l|l|l|l|}
\hline
\textbf{Source ID} &
  \textbf{RA, Dec (J2000)$^a$ } &
  \textbf{r ($^\prime$ )$^b$} &
  \textbf{Previous/New IDs} &
  \textbf{Period (days)} &
  \textbf{$\delta$ F336W (mag)} &
  \textbf{Mean F336W (mag)} &
  \textbf{$\mu_\alpha \cos (\delta)$ (mas/yr)$^c$} &
  \textbf{$\mu_\delta $ (mas/yr)$^d$} &
  \textbf{P (per cent)$^e$} \\ \hline
HUGS: 11228 &
  17:40:37.072 -53:40:08.99 &
  0.8 &
  \begin{tabular}[c]{@{}l@{}}HACKS: 10097\\ ACS: 11691\end{tabular} &
  $0.769 \pm 0.01$ &
  0.25 &
  19.65 &
  -0.11 &
  -0.57 &
  97.9 \\ \hline
\end{tabular}
}
\caption{\label{tab:table-name} $^a$ Coordinates from the HUGS catalogue \citep{Piotto2015AJ,Nardiello2018MNRAS}.$^b$ Distance from the centre of the cluster. $^c,^d$ data from the HACKS catalogue \citep{HACKSLibralato2022}. $^e$ Membership probability from the HUGS catalogue. }
\label{tab:table-detection}
\end{table*}

\section{Discussion}



\subsection{A Candidate He WD}

\begin{figure}
\includegraphics[width=1.0\columnwidth]{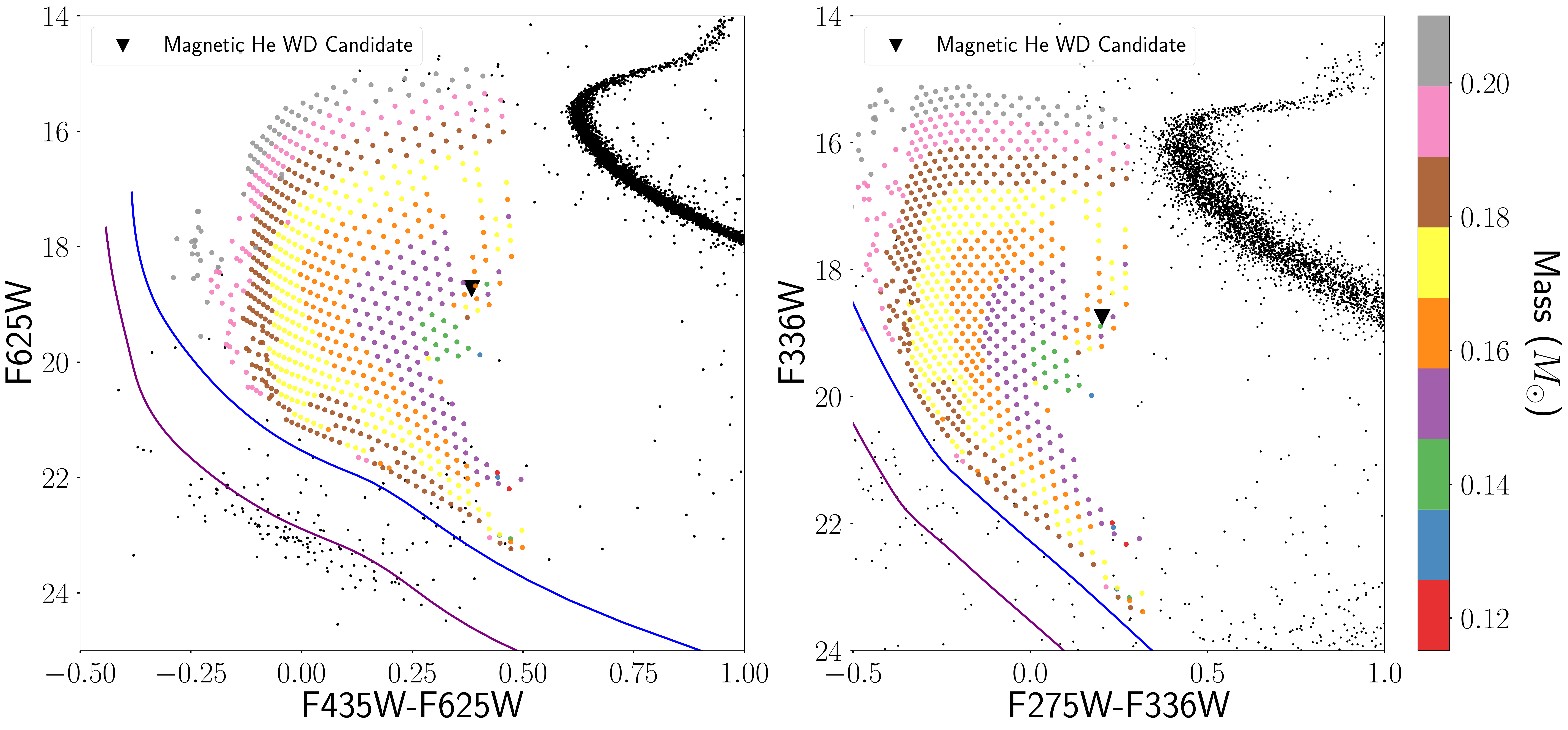}
\caption{Dereddened CMD of stars from the HACKS catalogue of NGC 6397. The black small circles represent stars in the clusters, and the black triangle indicates the position of the candidate magnetic helium-core white dwarf. The solid blue and purple lines represent the theoretical cooling sequence from \protect\cite{Althaus2009A&A...502..207A} for WDs of masses $M_{wd} = 0.22 M_\odot$ and $M_{wd} = 0.52 M_\odot$ respectively. The coloured circles are the theoretical evolutionary tracks from \protect\cite{Althaus2013ELMs} for WDs with $M_{wd} < 0.21 M_\odot$. The candidate magnetic He WD (black triangle) lies near the tracks predicted for a 0.16 $M_\odot$ He WDs, $0.24 R_{\odot}$ radius, an age of 0.11 Gyr, an effective temperature of 7586 K and surface gravity of $\log g = 4.9$}
\label{fig:ELMModels}
\end{figure}

Based on the position of the star in the CMDs and the lower limit on the radial velocity of this source, this system is a candidate He WD in a binary system. To estimate the mass and the age of the candidate He WD we use the theoretical evolutionary sequences, calculated for the metallicity of NGC~6397 ([Fe/H] = -2.03),  published by \cite{Althaus2013ELMs} as shown in fig. ~\ref{fig:ELMModels}. These authors used a statistical approach to create a grid of masses and ages for extremely low-mass He WDs. Such an approach is needed due to the fact that the different cooling tracks cross each other, and this leads to no unique solution for mass and age for a given $\log g$ and $T_{eff}$. To fit the available photometry for our target, we use model atmospheres described at length in \cite{Rohrmann2012A&A...546A.119R} and references therein. This model atmosphere code was also used to derive the outer boundary conditions for the mentioned evolving models. Synthetic magnitudes in the HST photometry are calculated using the zero-points derived from the Vega spectrum integrated over the passband for each filter. The input parameters of model atmospheres are the effective temperature, the surface gravity (evaluated from the stellar mass and radius of the evolutionary models), and the chemical composition, which we assume here to be pure hydrogen. In the F435W-F625W CMD, the candidate He WD (triangle) lies near the tracks predicted for a 0.16 $M_\odot$ He WD, with $log g = 4.9$, and  an age of 0.11 Gyr. As we showed in section~\ref{sec:lc} this is a variable object and we expect that the position on the CMD is phase-dependent and could introduce some bias in the derivation of the  exact physical parameters. The light curve (fig.~\ref{fig:F336WLC Folded}), shows clear variability of the order of $0.1$ magnitudes. The right panel of fig.~\ref{fig:ELMModels} shows the location of the candidate He WD on the F275W-F336W CMD assuming an average dereddened magnitude of  $m_{F336W}  = 18.54$.  The candidate He WD (triangle) lies near the tracks predicted for a 0.16 $M_\odot$ He WD,  $log g = 4.7$, and  an age of 0.068 Gyr. Assuming that the colour F275W-F336W stays the same, a 0.1 magnitude variability changes the estimate to a He WD with a mass of 0.16 $M_\odot$, $log g = 4.8$ and age of 0.088 Gyr. The differences in the estimated parameters do not change any of the conclusions of the paper, and all positions in the CMD are consistent with a He WD. For the rest of the paper, we assume a He WD with a mass of 0.16 $M_\odot$, $log g = 4.9$, and  an age of 0.11 Gyr.


\subsection{Can the periodic variability be explained by orbital modulation?}

We consider the possibility that the 0.76 day periodicity observed may be the orbital period of the system. Assuming it is the orbital period $P$, we can calculate the mass function for the companion
$$f(M) = \frac{P K_{wd}^3}{2 \pi \, G} = \frac{M_c^3 \sin ^3 i}{(M_c+M_{wd})^2} $$

where $K_{wd}$ is the semi-amplitude of the radial velocity for the WD, $M_c$ is the mass of the companion, $M_{wd}$ is the mass of the WD, and $q = M_{wd}/M_{c}$. Using the values for this source, an orbital period of 0.76 days and a radial velocity of $\sim 250$ km/s, this would give a lower limit for the companion of $\sim 1.23 M_\odot$ (assuming $M_{wd} = 0$ and inclination, i = 90\degr). At this orbital period and radial velocity, this value for the minimum companion mass already 
indicates 
that the 
companion is 
not a main sequence star, as this mass value is larger than the turnoff mass for the cluster ($\sim 0.8 M_\odot$)\citep{Richer2008AJ....135.2141R}.  The relation between the mass function of the companion, given above, and its actual mass, is $M=f(M)(1+q)^2 sin^{-3} i$.  For an exactly edge-on orbit, if the white dwarf mass is 0.16$M_\odot$ as given above, the mass of the companion would be $1.5 M_\odot$, allowing only a neutron star or black hole as a possible companion.  From the MUSE spectrum (fig.\ref{fig:MUSESpectrumFitBoth}), we measure a radial velocity of $\sim 250$ km/s.  Given that there was only a single measurement from MUSE, it is most likely that the $K_{wd}$ value for the white dwarf is actually substantially larger than the value obtained during that one measurement.  The 18 hour period is orders of magnitude longer than that at which a white dwarf can fill its Roche lobe, so no measurable ellipsoidal modulations would be expected if this were the orbital period, yielding an additional problem, that there is not a clear mechanism for producing photometric modulation on the observed timescale.


We arrive at a similar conclusion when we estimate an upper limit on the mass of the companion based on the reddest photometric  band available, F814W. In redder bands, a putative non-degenerate, main sequence, companion to the He WD would dominate. This is not the case, as seen in fig.~\ref{fig:F814WCMD}, where the object remains bluer than the main sequence in the redder bands. Assuming all the red light comes from the companion we can get  
a lower limit
for the absolute magnitude in F814W. Assuming an extinction of $A_{F814W} = 0.33$ towards the direction of the cluster  \citep{Richer2008AJ....135.2141R}, we get 
a lower limit 
for the absolute magnitude of $M_{F814W} = 6.44$. This value corresponds to a K6 star with a mass $~\sim 0.7 M_\odot$ \citep{Mamajek2013}. The mass upper limit from the brightness in the red band is lower than the lower limit from the mass function, making the companion underluminous for its mass and thus probably a second compact object.

Another possibility is that the orbital period is 1.52 days, twice the peak found in the periodogram, and due to ellipsoidal modulation of a stripped star, and implies we have a stripped star, instead of a WD, that is near Roche lobe filling. A star with colours $V_0 -I_0 = 0.29$, where $V_0$ and $I_0$ are the dereddened magnitudes in the F606W and F814W filters respectively, would have a $T_{eff}$ of 7400 K.  From this estimate for the $T_eff$, and the absolute magnitude, $M_{F814W} = 6.44$, calculated above, we can estimate a radius $R = 0.13 R_\odot$. Using the well-known density-period relation for a Roche lobe filling star ( $ \bar{\rho}_{mean} \approx 0.185 \,g \, cm^{-3} \,P_{orb}$),  this would imply a mass of $2 \times 10^{-4} M_\odot$, a mass comparable to a planet. This rules out the 
possibility 
that the periodicity is due to the ellipsoidal modulation of a stripped star.



Many He WDs are found in binary systems as companions to NSs, in either ultra-compact X-ray binaries  \citep[e.g.][]{Nelemans2006MNRAS.370..255N} or detached millisecond pulsar binaries \citep[e.g.][]{vanKerkwijk2010ApJ...715...51V,RiveraSandoval2015MNRAS.453.2707R}.
An ultra-compact X-ray binary can be immediately ruled out from the suggested orbital period, which is far too long to permit Roche-lobe overflow ($<$1 hour in ultra-compact X-ray binaries). 
%
He WDs in detached millisecond pulsar binaries can  show sinusoidal optical modulations, due to illumination of the companion by gamma-rays and/or pulsar winds from the pulsar, as in the MSP 47 Tuc U \citep{Edmonds01}. 
%
The lack of detected radio emission does not yet constrain the MSP hypothesis, as radio continuum observations \citep{Tudor22} do not approach the faintest MSPs yet known \citep{Wang21}.\footnote{Another, faint MSP was recently detected in NGC 6397 by \citealt{Zhang22}.}
 However, the lack of X-ray emission from this object indicates an MSP nature is unlikely. \citet{bogdanov_chandra_2010} placed an X-ray upper limit of $L_X$(0.5-6 keV)$=10^{29}$ erg/s (consistent with  \citealt{BahramiaArashn2020ApJ...901...57B}).
\citet{Lee18} and \citet{Zhao22} identify no MSPs (among 47  field MSPs, and 68 globular cluster MSPs, respectively) with $L_X$(0.5-6 keV)$<10^{29}$ erg/s; the least X-ray-luminous MSPs identified are PSR J0636+5129 ($L_X$(0.2-2 keV)$=1.3\times10^{29}$ erg/s, \citealt{Spiewak16}), 
PSR B1257+12 ($L_X$(0.3-8 keV)$=3\times10^{29}$ erg/s, \citealt{Pavlov07}) and PSR J0645+5158 ($L_X<3\times10^{29}$ erg/s, \citealt{Spiewak16}). This object would thus be the least X-ray luminous MSP yet known.  Additionally, given the wide separation, the companion star would be occupying a small solid angle from the point of view of the putative neutron star, meaning that it would be hard to obtain a 10\% flux modulation of the star via the heating from the faint, distant pulsar.

%
%


All this leads us to conclude that the modulation we see in the light curve is probably not due to the orbital period. Future spectroscopic and photometric follow-up can determine the orbital period and the nature of the binary companion of this system.

\subsection{A Candidate Magnetic He WD}

Another possibility we consider is that the periodic modulation is due to the magnetism of the WD. WDs with a high enough magnetic field ($>10^4$ G) can show modulation in the light curve on the order of $>0.1$ mag due to surface magnetic spots in the WD, 
via the process of magnetic dichroism \citep{1976ApJ...209..208L}. We  argue that this is what might be happening in this system and interpret the periodicity in the light curve as a periodic modulation due to a magnetic spot on a rotating He WD.

If the periodicity is due to the rotation of a magnetic white dwarf, we can estimate the initial rotational period of the system. Assuming that magnetic dipole radiation is solely responsible for the spin-down of the system, the evolution of the system can be written as $\dot \Omega = -K \Omega ^3$, where $\Omega$ and $\dot \Omega$ are the angular velocity and its derivative, respectively, and K is a proportionality constant. For a non-varying dipole magnetic field, the age (t) can be inferred from the equation above to be $    t = \tau \left [ 1-\left ( \frac{P_0}{P} \right )^2 \right ] $, where $\tau  \equiv \frac{P}{2 \dot P}$, and $\tau$ is the characteristic age, $P_0$ is the initial spin period, $P$ is the spin period, and $\dot P$ is the period derivative. In cgs units, assuming that the radiation luminosity is equal to the rotational energy loss rate, the period evolution can be written as $\dot P = \frac{8\pi^2}{3} \frac{B^2\times R^6}{c^3 I P} \sin^2 \alpha$, where $\alpha$ is the unknown inclination angle between the rotation and magnetic axes, R is the radius of the white dwarf, B is the magnitude of the magnetic field, and I the moment of inertia. 

Magnetic fields of isolated WDs are observed to lie in the range of $10^3-10^9$ G \citep{FerrarioMagneticWD}, while \cite{RotationMagWDBrinkworth2013ApJ...773...47B} shows that there are often measurable modulations due to spots in white dwarfs with magnetic fields as low as $10^5$ G. For a $10^9$ G magnetic field, assuming an angle of $90^\circ$ between the rotation and magnetic axes, the characteristic timescale would be 0.00025 Gyr (i.e. 250 kyr). This is much shorter than the estimated cooling age for the white dwarf (0.11 Gyr), obtained by comparing to  theoretical models. This suggests that the magnetic field is weaker to get a shorter $\dot P$. We can get an upper limit on the magnetic field of the He WD, assuming that the He WD started its life spinning very rapidly. To get a characteristic age of 0.11 Gyr, the magnetic field cannot be larger than $3.4 \times 10^{7} G$. This gives an estimated range for the magnetic field of this system of $B \sim 10^{4}$ to $3.4\times 10^{7} G$, with the lower end of the range possible only if the white dwarf was initially rotating much slower than its breakup velocity.

For this range of $B$ field, the linear or quadratic Zeeman effect could dominate the spectrum depending on the value of the B field. For values below $B <  10\times 10^{6} G$, the linear Zeeman effect produces a triplet pattern, and above $B > 10\times 10^{6} G$, the quadratic Zeeman effect dominates and causes a blue shift in the wavelength of all lines in the spectra  that varies as $\propto \lambda^2 n^4$ and can be approximated by $\Delta \lambda_q = -7.5 \times 10^{-23} \lambda^2 n^4 B^2$ \citep{QuadracticZeeman1939PhRv...55...52J,PrestonZeeman1970ApJ...160L.143P}. For H$\alpha$ (n=3) and H$\beta$ (n=4), for a $B = 5\times 10^6$ G the shifts correspond to  $-296$ km/s and $-696$ km/s. We do not see such a strong difference between H$\alpha$, and H$\beta$, suggesting that the shifts are due to orbital variations and not due to the Zeeman effect.  The small, statistically marginal discrepancy between the wavelengths for the two lines provides intriguing, but non-conclusive evidence that the quadratic Zeeman effect may be responsible for the differences, but better data would be required to make that measurement clearly and convincingly. 


\section{Conclusion}

We report a variable blue star in the globular cluster NGC~6397. The position in the CMDs and the spectrum is consistent with the star being a He WD in a binary system. The optical/near-UV light curve shows a periodicity at $\sim 18$ hours. We speculate that this periodicity is due to the rotation  of the WD, and possibly due to magnetic spots on the surface of the WD. This would make this object the first candidate magnetic He WD in any GC, the first candidate magnetic WD in a detached binary system in any GC and one of the few He WDs with a known rotation period. We propose that the large radial velocity is due to the WD being in a binary system with another compact object. Future observation of higher resolution optical or infrared spectra to see the Zeeman splitting could confirm the magnetic nature of this system, and follow-up photometry or phase-resolved spectroscopy could determine the orbital period for this system and help identify the companion for this magnetic He WD candidate.

\section*{Acknowledgements}
We thank Thomas Kupfer and Alekzander Kosakowski for useful discussions. DB acknowledges financial support from FONDECYT (grant number 3220167). CH acknowledges support from NSERC, grant RGPIN-2016-04602. MPM also would like to thank David Zurek for useful discussions. MPM acknowledges financial support from the Kalbfleisch Fellowship, Richard Gilder Graduate School, American Museum of Natural History
\emph{Software used:} PyAstronomy \citep{pyastronomypackage}, SciPy \citep{SciPy-NMeth}, Astropy \citep{astropy:2013,astropy:2018}, Matplotlib \citep{matplotlib}, specutils \citep{specutilnicholas_earl_2021_4603801}.

\section*{Data Availability}

This work used public {\it HST} data available in the Mikulski Archive for Space Telescopes under the programs GO-10424, GO-10257, GO-10775 and GO-13297. The MUSE data can be found at \url{http://muse-vlt.eu/science}, and is based on observations obtained at the Very Large Telescope (VLT) of the European Southern Observatory, Paranal, Chile (ESO Programme ID 60.A-9100(C)).



\bibliographystyle{mnras}
\bibliography{example} 

\bsp	
\label{lastpage}
\end{document}